\documentclass[preprint,showpacs]{revtex4}

\usepackage{graphicx}
\usepackage{dcolumn}
\usepackage{bm}

\begin{document}
\bibliographystyle{prsty}
\newcommand{\bq}{\begin{mathletters}}
\newcommand{\eq}{\end{mathletters}}
\newcommand{\beq}{\begin{eqnarray}}
\newcommand{\eeq}{\end{eqnarray}}
\newcommand{\beqq}{\begin{eqnarray*}}
\newcommand{\eeqq}{\end{eqnarray*}}

\newcommand{\st}[1]{{\mbox{${\mbox{\scriptsize #1}}$}}}  
\newcommand{\BM}[1]{\mbox{\boldmath $#1$}}              
\newcommand{\uv}[1]{\mbox{$\widehat{\mbox{\boldmath $#1$}}$}}   
\newcommand{\dy}[1]{\mbox{\boldmath $\overline{#1}$}}   
\newcommand{\sbtex}[1]{{\mbox{\scriptsize #1}}}   
\newcommand{\nab}{\mbox{\boldmath $\nabla$}}            
\newcommand{\CS}{\mbox{$\begin{array}{c}\cos \\ \sin \\ \end{array}$}}
\newcommand{\SC}{\mbox{$\begin{array}{c}\sin \\ \cos \\ \end{array}$}}

\newcommand{\rmi}{{\rm i}}
\newcommand{\er}{{\bf e}_{r}}
\newcommand{\ep}{{\bf e}_\varphi}
\newcommand{\et}{{\bf e}_\theta}
\newcommand{\hh}{{\bf H}}
\newcommand{\ee}{{\bf E}}
\newcommand{\ww}{{\bf W}}
\newcommand{\vv}{{\bf v}}
\newcommand{\rr}{{\bf r}}
\newcommand{\eee}{{\bf e}}
\newcommand{\uu}{{\bf u}}
\newcommand{\lom}{{\bf L}}

\newcommand{\rmq}{{\rm q}}
\newcommand{\rmt}{{\rm t}}
\newcommand{\co}{{\rm co}}
\newcommand{\cl}{{\rm cl}}

\newcommand{\epr}{\varepsilon_{r r}}
\newcommand{\mur}{\mu_{r r}}
\newcommand{\alr}{\alpha_{r r}}
\newcommand{\kar}{\kappa_{r r}}
\newcommand{\Om}{\Omega_a^r}
\newcommand{\ve}{\vec{e}}
\newcommand{\vw}{\vec{w}}
\newcommand{\va}{\vec{a}}


\title{Electromagnetic Interaction of Arbitrary Radial-Dependent Anisotropic Spheres and Improved Invisibility for Nonlinear- Transformation-Based Cloaks}
\author{Cheng-Wei Qiu$^{1,2}$, Andrey Novitsky$^{3}$, Hua Ma$^{4}$, and Shaobo Qu$^{4}$}
\affiliation{$^{1}$Research Laboratory of Electronics, Massachusetts Institute of Technology, 77 Massachusetts Anvenue, Cambridge, MA 02139, USA.}\email{cwq@mit.edu}
\affiliation{$^{2}$Department of Electrical and Computer Engineering, National
University of Singapore, Kent Ridge, Singapore 119620.}
\affiliation{$^{3}$Department of Theoretical Physics, Belarusian State University, Nezavisimosti Avenue 4, 220050 Minsk, Belarus.}
\affiliation{$^{4}$The College of Science, Air Force University of Engineering, Xi'an 710051, China.}

\date{\today}

\begin{abstract}
An analytical method of electromagnetic wave interactions with a
general radially anisotropic cloak is established. It is able to
deal with arbitrary parameters ($\varepsilon_r(r)$, $\mu_r(r)$,
$\varepsilon_t(r)$ and $\mu_t(r)$) of a radially anisotropic
inhomogeneous shell. The general cloaking condition is proposed from
the wave relations for the first time. We derive the parameters of a
novel class of spherical nonlinear cloaks and examine its
invisibility performance by the proposed method at various nonlinear
situations. Spherical metamaterial cloaks with improved
invisibility performance is achieved with optimal nonlinearity in transformation and core-shell ratio.

\end{abstract}

\maketitle

\section{Introduction}

Coordinate transformation \cite{Andrey_1,Andrey_2,Andrey_3,Andrey_4}
for the design process of the cloaking devices has received great
attention. The cylindrical/spherical cloaking idea proposed by
Pendry  \cite{Andrey_1} is to employ radial anisotropic materials
whose parameters are determined from the topological variation
between the original and transformed spaces, based on the invariance
of Maxwell's equations throughout a specific coordinate
transformation \cite{Andrey_3}. The idea of cylindrical cloaking was
confirmed by analytical/full-wave methods
\cite{Andrey_5,Andrey_6,Andrey_7} and verified by an experiment
using artificial metamaterials with inclusions of metallic
split-ring resonators (SRRs) \cite{Andrey_8}. So far, significant
progress has been made on the study of cylindrical invisibility
cloaks. It reveals that the simplified parameters for cylindrical
cloaking still allow wave interactions with the cloaked object
\cite{Andrey_9} and the invisibility performance of a cylindrical
cloak is very sensitive to the geometrical perturbation of its
interior boundary \cite{Yan_JOSAA}, which can be both fixed by
introducing PEC/PMC linings onto the inner surface of the shell
\cite{Andrey_4,Andrey_10}. Since it is challenging to synthesize the
magnetic response in optical regime, nonmagnetic cylindrical cloaks
have been proposed by using quadratic transformation \cite{New_1}
and the general high-order transformation for nonmagnetic
cylindrical cloaks in optical frequency is addressed more recently
\cite{New_2}. Nevertheless, it is still difficult to realize the
position-dependent cylindrical cloak due to the limited resource of
natural materials exhibiting radial anisotropy \cite{Qiu_JOSA}. In
view of this, Cai {\itshape et al.} proposed a multilayered
cylindrical cloak by dividing the original position-dependent cloak
into many thin coatings in which the material parameters become
homogeneous \cite{Cai_Dehesa}. Furthermore, the cylindrical cloak
has been theoretically realized by a concentric cylinder of
isotropic homogeneous multilayers \cite{OptExpress}.
Arbitrary-shaped 2D cloaks have been investigated theoretically and
numerically \cite{Andrey_11,theoretical1,numerical1}.

However, for spherical invisibility cloaks there are still a lot of
unknowns to be explored because of the complexity in analysis and
simulation of scattering properties. Anisotropic and
position-dependent ideal spherical cloak based on the linear
transformation was suggested by Pendry \cite{Andrey_1}, and it has
been shown that spherical cloaks are less sensitive to the
perturbation than cylindrical cloaks \cite{Yan_JOSAA}, which is
mathematically proved \cite{Greenleaf}. There are several main
streams of studying linear first-order spherical invisibility
cloaks, whose the required materials and the corresponding methods
are distinct. The first approach is the classic cloak
\cite{Andrey_1}, which is linear, anisotropic and inhomogeneous. In
this connection, explicitly electromagnetic fields have been
formulated \cite{Andrey_12} and it is further confirmed that the
wave cannot interact with the concealed object \cite{Andrey_13}. The
second is to utilize a homogeneous anisotropic metamaterial cover to
achieve electromagnetic invisibility \cite{Andrey_14} via the
core-shell system. The third is the implementation of isotropic
plasmonic materials as the cloak based on cancellation scheme
\cite{Alu1,Alu2}. The fourth is to substitute the Pendry's classic
cloak with alternating thin multi-shells and each shell is
homogeneous and isotropic \cite{SphericalCloaking}. Each approach
mentioned above has its own advantages and restrictions. For
instance, the first approach \cite{Andrey_1} requires higher
complexity in material parameters, and the analysis is situated
towards a particular anisotropy ratio, which is addressed in
\cite{SphericalCloaking}. The second approach \cite{Andrey_14}
removes the requirement of material inhomogeneity, in which
parameters are position-independent. However its cloaking property
is quite reliant on the core-shell ratio, and the same feature is
possessed by the third approach (cancellation scheme). The fourth
method has less restrictions on materials but needs a lot of
coatings which are sufficiently thin compared with the wavelength.
The high-order term in the refractive index of an inhomogeneous
spherical lens is discussed and its possibility of realizing a
spherical cloak without parametric singularity is addressed
\cite{New_3}. The critical material singularity is thus transformed
into the geometrical singularity which is less demanding
\cite{Andrey_15}.

In this paper, a more general high-order nonlinear transformation will be considered for spherical cloaks. We first propose a general algorithm to study the
electromagnetic scattering by a particle coated by a radially inhomogeneous
shell whose anisotropic parameters can be arbitrary. We discretize the shell into multiple spherical shells, each of which is homogeneous and anisotropic. Also, we propose a novel class of nonlinear transformation based
spherical cloaks, whose anisotropy ratio is position-dependent and
also too complicated to be treated by any mentioned methods. By
utilizing the established general scattering algorithm, the
invisibility performance and its dependence on the nonlinear
transformation are investigated. Finally, the numerical results
suggest a particular type of nonlinear spherical cloak providing
better invisibility than Pendry's linear spherical cloak.

\section{Scattering Algorithm for a General Radially Aniostropic Metamaterial Cloak}

\begin{figure}[htbp]
\centering\includegraphics[width=6.5cm]{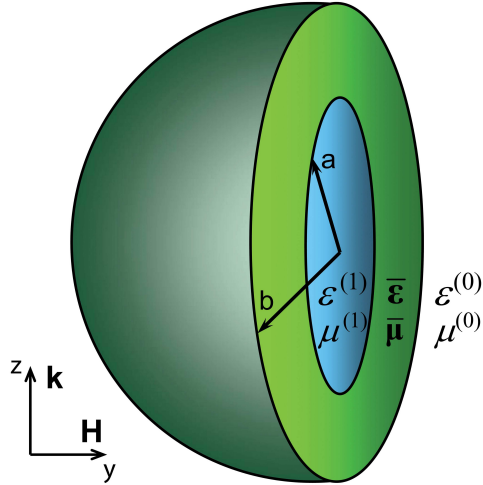}
\caption{The geometry of the spherical cloak structure. Incident
plane wave is propagating along z-direction and its electric field
is polarized along x-direction.  The supscripts $0$ and $1$ denote
the parameters of the host and cloaked media, respectively. The
anisotropic $\dy{\varepsilon}$ and $\dy{\mu}$ presents the
parameters of the cloak shell.}\label{Geometry_Andrey}
\end{figure}

Fig.~\ref{Geometry_Andrey} illustrates the configuration of the
cloak structure, i.e. the inner and outer radii are denoted by $a$
and $b$ respectively; innermost region is filled by an isotropic
dielectric material characterized by $\varepsilon^{(1)}$ and
$\mu^{(1)}$; intermediate region is occupied by a general spherical
metamaterial cloak characterized by $\dy{\varepsilon}$ and
$\dy{\mu}$
\begin{eqnarray}
\dy{\varepsilon} = \varepsilon_r(r) \er \otimes \er + \varepsilon_t(r) \dy{I}_t, \qquad
\dy{\mu} = \mu_r(r) \er \otimes \er + \mu_t(r) \dy{I}_t, \label{eps_mu}
\end{eqnarray}
where $\varepsilon_r$ and $\mu_r$ are the radial
permittivity and permeability, $\varepsilon_t$ and $\mu_t$
are the transversal material parameters, $\dy{I}_t = \dy{I} - \er \otimes
\er = \et \otimes \et + \ep \otimes \ep$ is the projection operator
onto the plane perpendicular to the vector $\er$, $\dy{I}$ is the
unit three-dimensional dyad, unit vectors $\er$, $\et$, and $\ep$
are the basis vectors of the spherical coordinates.

In this section, the scattering theory of multilayer anisotropic
spherical particles is provided and applied to study a cloak. We
suppose that the arbitrary field distribution of the incident
monochromatic wave interacts with the two-layer sphere.

Using the separation of variables, the solution of Maxwell's
equations in spherical coordinates ($r$, $\theta$, $\varphi$) can be
presented as
\begin{eqnarray}
{\bf E} (r, \theta, \varphi) = F_{lm}(\theta, \varphi) {\bf E}(r), \qquad
{\bf H} (r, \theta, \varphi) = F_{lm}(\theta, \varphi) {\bf H}(r),
\label{solution1}
\end{eqnarray}
where the designation $\ee (r)$ means that the components of the
electric field vector depend only on the radial coordinate $r$ as
$E_r(r)$, $E_\theta(r)$, and $E_\varphi(r)$ (however, the vector
itself includes the angle dependence in the basis vectors), and the
second rank tensor in three-dimensional space $F_{lm}$ serves to
separate the variables ($l$ and $m$ are the integer numbers). It can
be written as the sum of dyads:
\begin{equation}
F_{lm}=Y_{lm} \er \otimes \er + {\bf X}_{lm} \otimes \et + (\er
\times {\bf X}_{lm}) \otimes \ep. \label{Flm}
\end{equation}
where $Y_{lm}(\theta, \varphi)$ and ${\bf X}_{lm}(\theta, \varphi)$
are the scalar and vector spherical harmonics, the orthogonality of which has been well described in \cite{Jackson}. Tensor functions $F_{lm}$ are very useful, because they completely describe the angle
dependence of the spherical electromagnetic waves and satisfy the
orthogonality conditions
\begin{equation}
\int_0^\pi \int_0^{2 \pi} F_{l'm'}^+(\theta, \varphi) F_{lm}(\theta,
\varphi) \sin \theta {\rm d} \theta {\rm d} \varphi = \dy{I}
\delta_{l'l} \delta_{m'm}, \label{orthogF}
\end{equation}
where the superscript $+$ stands for the Hermitian conjugate.

From the commutation of $\dy{\varepsilon}$, $\dy{\mu}$ and $F_{lm}$, it follows
that the electric and magnetic fields obey the set of ordinary
differential equations
\begin{eqnarray}
&&\er^\times \frac{{\rm d} \hh}{{\rm d} r} + \frac{1}{r} \er^\times
\hh - \frac{\rmi \sqrt{l(l+1)}}{r} \ep^\times \hh = - \rmi k_0
\dy{\varepsilon}\cdot \ee, \nonumber \\
&&\er^\times \frac{{\rm d} \ee}{{\rm d} r} + \frac{1}{r} \er^\times
\ee - \frac{\rmi \sqrt{l(l+1)}}{r} \ep^\times \ee = \rmi k_0
\dy{\mu}\cdot \hh, \label{ODE1}
\end{eqnarray}
where $k_0 = \omega / c$ is the wavenumber in vacuum, and $\omega$
denotes the circular frequency of the incident electromagnetic wave.
Quantity ${\bf n}^\times$ is called the tensor dual to the vector ${\bf
n}$ \cite{Fedorov}. It results in the vector
product, if multiplied by a vector ${\bf a}$ as ${\bf n}^\times {\bf
a} = {\bf n} {\bf a}^\times = {\bf n} \times {\bf a}$.

Eq.~(\ref{ODE1}) results from variable separation in
Maxwell's equations, and includes two algebraic scalar
equations, therefore, two field components, $H_r$ and $E_r$, can be
expressed by means of the rest four components. This can be
presented in terms of the matrix link between the total fields
$\hh=\hh_\rmt+H_r \er$ and $\ee=\ee_\rmt+E_r \er$ and their
tangential components $\hh_\rmt$ and $\ee_\rmt$:
\begin{eqnarray}
\left( \begin{array}{c} \hh (r) \\ \ee (r) \end{array} \right) = V
(r) \left( \begin{array}{c} \hh_\rmt (r) \\ \ee_\rmt (r) \end{array}
\right), \qquad &V(r) = \left( \begin{array}{cc} \dy{I} & \frac{
\sqrt{l (l+1)}}{\mu_{r}(r) k_0 r} \er \otimes \et \\
- \frac{\sqrt{l (l+1)}}{\varepsilon_{r}(r) k_0 r} \er \otimes \et &
\dy{I}
\end{array} \right). \label{restore}
\end{eqnarray}
Excluding the radial components of the fields from Eq.~(\ref{ODE1}),
we arrive at a set of ordinary differential equations of the first
order for the tangential components, which can be interconnected by a four-dimensional
vector $\ww(r)$ as
\begin{equation}
\frac{{\rm d} \ww(r)}{{\rm d} r}=\rmi k_0 M(r)\ww(r), \label{Meqt}
\end{equation}
where
\[ M=\left(\begin{array}{cc}A&B\\C&D\end{array}\right), \qquad
\ww=\left(\begin{array}{c}\hh_\rmt\\ \ee_\rmt\end{array}\right)
\equiv \left(\begin{array}{c} H_\theta \\ H_\varphi \\ E_\theta \\
E_\varphi \end{array}\right) ,
\]
\begin{eqnarray}
A = D = \frac{\rmi}{k_0 r} \dy{I}, ~~
B=\varepsilon_t(r) \er^\times - \frac{l (l+1)}{\mu_r(r) k_0^2 r^2} \ep
\otimes \et, ~~
C= - \mu_t(r) \er^\times + \frac{l (l+1)}{\varepsilon_{r}(r) k_0^2 r^2}
\ep \otimes \et. \label{ABCD}
\end{eqnarray}

Since $\hh_\rmt$ and $\ee_\rmt$ are continuous at the spherical interface,
they can be used for solving the scattering problem. Now, we analyze the situation of $r$-dependent permittivities and
permeabilities, which arise from the spherical cloaking. Excluding the $\varphi$-components of fields from Eq.~({\ref{Meqt}), we derive the differential equation of the second order for the
vector $w_\theta = \et \cdot \ww = (H_\theta, E_\theta)$:
\begin{equation}
w''_\theta + \frac{2}{r} w'_\theta - \left(\begin{array}{cc}
\frac{\varepsilon'_t}{\varepsilon_t} & 0 \\ 0 & \frac{\mu'_t}{\mu_t}
\end{array}\right) w'_\theta + \left[ k_0^2 \varepsilon_t \mu_t -
\frac{1}{r} \left(\begin{array}{cc}
\frac{\varepsilon'_t}{\varepsilon_t} & 0 \\ 0 &
\frac{\mu'_t}{\mu_t}\end{array}\right) - \frac{l (l+1)}{r^2}
\left(\begin{array}{cc} \frac{\varepsilon_t}{\varepsilon_r} & 0 \\ 0
& \frac{\mu_t}{\mu_r}
\end{array}\right) \right] w_\theta = 0, \label{wThetaInhom}
\end{equation}
where the prime denotes the derivative with respect to  $r$. Further we will apply
one condition on the medium parameters, which is usually used for
the spherical cloaks: $\varepsilon_r(r) = \mu_r(r)$ and $\varepsilon_t(r) =
\mu_t(r)$ due to the impedance matching. Then the equations for $H_\theta$ and $E_\theta$ coincide
and can be written in the form
\begin{equation}
w''_\theta + \left( \frac{2}{r} -
\frac{\varepsilon'_t}{\varepsilon_t} \right) w'_\theta + \left[
k_0^2 \varepsilon_t^2 - \frac{\varepsilon'_t}{r \varepsilon_t} -
\frac{l (l+1)}{r^2} \frac{\varepsilon_t}{\varepsilon_r} \right]
w_\theta = 0. \label{wThetaInhomCl}
\end{equation}

This equation can be solved analytically only in very few cases. As
an example, we can offer a case of
$\varepsilon_t =\mu_t=  a_1/r$ and $\varepsilon_r =\mu_r =a_2/r$, where $a_{1,2}$ are arbitrary values. However such dependencies do not provide the cloak
properties. Another solvable case in Eq.~(\ref{wThetaInhomCl}) is just Pendry's cloak, that is, $\varepsilon_t =\mu_t=b/(b-a)$ and $\varepsilon_r =\mu_r
\varepsilon_t (r - a)^2/r^2$.

Although analytical solutions cannot be found for all situations,
the general structure of solutions can be studied. The solution of
two differential equations of the second order
Eq.~(\ref{wThetaInhom}) contains four integration constants $c_1$,
$c_2$, $c'_1$, and $c'_2$. The constants can be joined together into
a couple of vectors ${\bf c}_1 = c_1 \et + c'_1 \ep$ and ${\bf c}_2
= c_2 \et + c'_2 \ep$. $\varphi$-components of the field vectors
$H_\varphi$ and $E_\varphi$ are expressed in terms of the
$\theta$-components which have been already determined. The relation
between $\theta$- and $\varphi$-components follows from
Eq.~(\ref{Meqt}). Summing up both components, the resultant field
can be presented as
\begin{equation}
\ww=S(r){\bf C}, \quad S(r)=\left( \begin{array}{cc} \eta_1(r) &
\eta_2(r) \\ \zeta_1(r) & \zeta_2(r) \end{array} \right), \quad {\bf
C}= \left( \begin{array}{c} {\bf c}_1 \\ {\bf c}_2 \end{array}
\right), \label{WSC}
\end{equation}
where $\eta_{1,2}$ and $\zeta_{1,2}$ are the
two-dimensional blocks of the matrix $S(r)$. \{$\eta_1$,~$\zeta_1$,~${\bf c}_1$\} and \{$\eta_2$,~$\zeta_2$,~${\bf c}_2$\} denote the first and second sets of the independent solution of Eq.~(\ref{wThetaInhom}), respectively. Therefore, the general solution can be decomposed into two
terms as $\ww = \ww^{(1)} + \ww^{(2)}$, where
\begin{eqnarray}
\ww^{(1)}= \left( \begin{array}{cc} \hh_{\rmt 1} \\
\ee_{\rmt 1} \end{array} \right) = \left( \begin{array}{cc} \eta_1  \\
\zeta_1 \end{array} \right) {\bf c}_1, \qquad
\ww^{(2)}= \left( \begin{array}{cc} \hh_{\rmt 2} \\
\ee_{\rmt 2} \end{array} \right) = \left(
\begin{array}{cc} \eta_2  \\ \zeta_2 \end{array} \right) {\bf c}_2.
\label{separatewaves}
\end{eqnarray}

Electric and magnetic fields of each independent wave are related
by means of impedance tensor $\Gamma$ as $\ee_{\rmt j}=\Gamma_j
\hh_{\rmt j}$ ($j=1,2$). Thus, the impedance tensor equals
\begin{equation}
\Gamma_j(r)=\zeta_j(r) \eta^{-1}_j(r). \label{gen_Impedance}
\end{equation}

Vectors ${\bf c}_1$ and ${\bf c}_2$ can be expressed by means of the
known tangential electromagnetic field $\ww(a)$ as ${\bf C} =
S^{-1}(a)\ww (a)$. Thus Eq.~(\ref{WSC}) can be rewritten as follows
\begin{equation}
\ww(r)=\Om \ww(a), \qquad \Om=S(r) S^{-1}(a), \label{gen_evol_oper}
\end{equation}
where the evolution operator (transfer matrix) $\Om$ connects
tangential field components at two distinct spatial points, i.e.,
$r$ and $a$. One can obtain the complete solution of the fields
$\ee(\rr)$ and $\hh(\rr)$ by summing over $l$ and $m$ in the subsequent tensor
product of $F_{lm}(\theta, \varphi)$ describing angle dependence
(Eq.~(\ref{Flm})), matrix $V^l(r)$ restoring the fields with their
tangential components (Eq.~(\ref{restore})), and tangential field
vectors (Eq.~(\ref{WSC})):
\begin{eqnarray}
\left( \begin{array}{cc} \hh (\rr)
\\ \ee(\rr) \end{array} \right)= \sum_{l=0}^\infty \sum_{m=-l}^l
\left( \begin{array}{cc} F_{lm} (\theta, \varphi) & 0
\\ 0 & F_{lm} (\theta, \varphi) \end{array} \right)
V^l(r) \left( \begin{array}{cc} \eta^l_1(r) & \eta^l_2(r)
\\ \zeta^l_1(r) & \zeta^l_2(r) \end{array} \right) \left(
\begin{array}{c} {\bf c}^{lm}_1 \\ {\bf c}^{lm}_2 \end{array}
\right). \label{gen_sol}
\end{eqnarray}

In general, the solutions cannot be studied in the closed form for
nonlinear spherical cloaks. Therefore, the approximate method of
numerical computations is applied. An inhomogeneous anisotropic
spherical cloak $a<r<b$ is equally divided into $N$ homogeneous anisotropic
spherical layers, i.e., replaced with a multi-layer structure. The
number of the layers strongly determines the accuracy of
calculations. The $j$-th homogeneous shell is situated in the region
between $r=a_{j-1}$ and $r=a_j$, where $j=1,\ldots,N$, $a_{0}=a$ and
$a_{N}=b$. Wave solution of the single homogeneous layer can be
represented in the form of the evolution operator
$\Omega_{a_{j-1}}^{a_j}$. The solution for the whole inhomogeneous
shell is thus the subsequent product of the elementary evolution
operators:
\begin{equation}
\Omega_{a}^b =  \Omega_{a_{N-1}}^b \ldots \Omega_{a_{1}}^{a_2}
\Omega_{a}^{a_1}. \label{evolut_layers}
\end{equation}

The solution of Eq.~(\ref{wThetaInhom}) in one layer with constant
permittivities $\varepsilon_r$, $\varepsilon_t$ and permeabilities
$\mu_r$, $\mu_t$ is expressed by means of a couple of
independent spherical functions $g^{(1)}_{\nu}$ and $g^{(2)}_{\nu}$:
\begin{eqnarray}
\left( \begin{array}{c} H_\theta (r)\\ E_\theta (r) \end{array}
\right)
=\left( \begin{array}{c} g^{(1)}_{\nu_1}(k_t r) c_1 +
g^{(2)}_{\nu_1}(k_t r) c_2
\\ g^{(1)}_{\nu_2}(k_t r) c'_1 + g^{(2)}_{\nu_2}(k_t r)
c'_2 \end{array} \right), \label{theta-solution}
\end{eqnarray}
where $k_t = k_0 \sqrt{\varepsilon_t \mu_t}$, $\nu_1 = \sqrt{l (l+1)
\varepsilon_t/\varepsilon_r + 1/4} - 1/2$, $\nu_2 = \sqrt{l (l+1)
\mu_t/\mu_r + 1/4} - 1/2$ which applies to both uniaxial anisotropic and bianisotropic media \cite{Andrey_15,Andrey_16,Novitsky08}. Functions $g^{(1,2)}_{\nu}$ of the order
$\nu$ can be spherical Bessel functions, modified spherical Bessel
functions, or spherical Hankel functions.

Blocks $\eta$ and $\zeta$ introduced in Eq.~(\ref{WSC}) are the tensors
\begin{eqnarray}
&& \eta_{1,2}= g^{(1,2)}_{\nu_1} \et \otimes \et
- \frac{\rmi} {\mu_t k_0 r} \frac{{\rm d}(r
g^{(1,2)}_{\nu_2})}{{\rm d} r}  \ep \otimes \ep, \nonumber \\
&& \zeta_{1,2}= g^{(1,2)}_{\nu_2} \et \otimes \ep
+ \frac{\rmi} {\varepsilon_t k_0 r} \frac{{\rm d}(r
g^{(1,2)}_{\nu_1})}{{\rm d} r} \ep \otimes \et, \label{etazeta_ex}
\end{eqnarray}
which can be presented as two-dimensional matrices for computation
purposes.

Now we turn to the scattering of electromagnetic waves from the
cloaking structure depicted in Fig.~\ref{Geometry_Andrey}. We
suppose that an electromagnetic field $\hh_{\rm inc}(\rr)$ and
$\ee_{\rm inc}(\rr)$ is incident on the coated spherical particle
from air ($\varepsilon^{(0)}=1$, $\mu^{(0)}=1$). Wave solutions in each of the $N$ layers can be written using the
general solution Eq.~(\ref{gen_sol}), which is already known.
Scattered field propagating in air can be presented by the
superposition of diverging spherical waves which are mathematically
described by spherical Hankel functions of the first kind
$h^{(1)}_\nu(x)$. Let us first introduce $\widetilde{\eta}$ and
$\widetilde{\zeta}$ which correspond to the tensors $\eta$ and
$\zeta$ in Eq.~(\ref{etazeta_ex}) when Hankel functions replace
Bessel functions. Then we obtain the scattered fields
\begin{eqnarray}
\left( \begin{array}{cc} \hh_{\rm sc} (\rr)
\\ \ee_{\rm sc}(\rr) \end{array} \right)= \sum_{l=0}^\infty \sum_{m=-l}^l \left(
\begin{array}{cc} F_{lm} & 0 \\ 0& F_{lm}
\end{array} \right)
V^l_{\rm sc}(r) \left(
\begin{array}{c} I \\ \widetilde{\Gamma}^l(r) \end{array} \right) \widetilde{\eta}^l(r)
(\widetilde{\eta}^l(b))^{-1} \hh^{lm}_{\rm sc}(b),
\label{scat_field}
\end{eqnarray}
where $\widetilde{\Gamma}^l=\widetilde{\zeta}^l
(\widetilde{\eta}^l)^{-1}$ is the impedance tensor of the $l$th
scattered wave, and $\hh^{lm}_{\rm sc}(b)$ is the tangential
magnetic field at the outer interface $r=b$. Applying the evolution
operator $\Omega_{a}^r$, the electromagnetic field in the shell
takes the form
\begin{eqnarray}
\left( \begin{array}{cc} \hh_{{\rm sh}} (\rr)
\\ \ee_{{\rm sh}}(\rr) \end{array} \right) = \sum_{l=0}^\infty \sum_{m=-l}^l
\left( \begin{array}{cc} F_{lm} & 0 \\ 0& F_{lm}
\end{array} \right)
V^l_{{\rm sh}}(r) \Omega_{a}^r \left(
\begin{array}{c} I \\ \Gamma^l_1(a)
\end{array} \right) \hh^{lm}_{1}(a),
\label{inside_field}
\end{eqnarray}
where $\Gamma^l_1=\zeta^l_1 (\eta^l_1)^{-1}$ is the impedance tensor
of the $l$th wave inside the inner sphere (region 1), and
$\hh^{lm}_{1}(a)$ is the tangential magnetic field at the inner
interface of the shell $r=a$.

By projecting the fields onto the outer interface $r=b$ and
integrating over the angles $\theta$ and $\varphi$ with the help of
orthogonality condition Eq.~(\ref{orthogF}), we derive the boundary
conditions
\begin{eqnarray}
\ww_{\rm inc}^{lm}  +  \left( \begin{array}{c} I \\
\widetilde{\Gamma}^l(b) \end{array} \right) \! \hh^{lm}_{\rm sc}(b)
= \Omega_{a}^{b} \!\! \left( \begin{array}{c} I \\
\Gamma^l_1(a) \end{array} \right) \! \hh^{lm}_{1}(a),
\label{bc_final}
\end{eqnarray}
where
\begin{equation}
\ww_{\rm inc}^{lm} = \int_0^\pi \int_0^{2 \pi} \left(
\begin{array}{cc} F^+_{lm} (\theta, \varphi) I \hh_{\rm
inc} (b,\theta, \varphi)
\\ F^+_{lm} (\theta,\varphi) I \ee_{\rm
inc}(b,\theta, \varphi) \end{array} \right) \sin \theta {\rm d}
\theta {\rm d} \varphi. \label{Wlm}
\end{equation}

Eq.~(\ref{bc_final}) represents the system of four linear equations
for four components of the vectors $\hh_{\rm sc}^{lm}$ and
$\hh_{1}^{lm}$. Finally, one can derive the amplitude of the
scattered electromagnetic field (see Appendix)
\begin{eqnarray}
\hh^{lm}_{\rm sc} (b) = - \left[ \left( \begin{array}{cc}
\Gamma^l_1(a) & - I \end{array} \right) \Omega_{b}^{a} \left(
\begin{array}{c} I \\ \widetilde{\Gamma}^l(b)
\end{array} \right) \right]^{-1}
\left[ \left( \begin{array}{cc} \Gamma^l_1(a) & - I
\end{array} \right) \Omega_{b}^{a} \ww_{\rm inc}^{lm}
\right]. \label{c_sc}
\end{eqnarray}

The scattered field in far zone can be characterized by the differential
cross-section (power radiated to $\er$-direction per solid angle
${\rm d}o$)
\begin{equation}
\frac{{\rm d} \sigma}{{\rm d} o}=r^2 \frac{|\hh_{\rm
sc}(\rr)|^2}{|\hh_{\rm inc}(\rr)|^2}.
\end{equation} In our notations, the differential cross-section averaged over the
azimuthal angle $\varphi$ (over polarizations) normalized by the
geometrical cross-section $\sigma_g=\pi b^2$ takes the form
\begin{equation}
\frac{{\rm d} \sigma}{\sigma_g \sin\theta {\rm d} \theta}=
\frac{1}{\sigma_g |\hh_{\rm inc}|^2} \sum_{m=-\infty}^\infty \left|
\sum_{l=|m|}^\infty \rmi^{-l-1} F_{lm} (\theta, 0)
\widetilde{\eta}_l^{-1}(b) \hh^{lm}_{\rm sc}(b) \right|^2.
\label{cross}
\end{equation}

From the point of view of the scattering theory, it is
straightforward to define a cloak as one specially matched layer
that provides zero scattering for arbitrary materials inside. In \cite{Andrey_12} zero scattering was proved analytically for the Pendry's spherical cloak. Here, we have proposed a more general scattering
algorithm for radially anisotropic materials, which is useful in
studying the scattering of spherical cloaks based on complex (e.g.,
high-order, nonlinear, etc.) transformations. From the proposed
scattering theorem, we can determine the invisibility condition
(zero scattering) specified by the condition $\hh^{lm}_{\rm
sc}(b)=0$, which in turn can be rewritten using Eq.~(\ref{c_sc}) as
follows
\begin{eqnarray}
\left( \begin{array}{cc} \Gamma^l_1(a) & - I
\end{array} \right) \Omega_{b}^{a} \ww_{\rm inc}^{lm}
=0. \label{hsc=0}
\end{eqnarray}
Arbitrary incident electromagnetic field $\ww_{\rm inc}^{lm}$ can be
excluded from this expression. In fact, zero scattering can be obtained for the trivial situation: electromagnetic field is
scattered by a ``virtual" air sphere at radius $b$. This assumption can be
presented in the form analogous to Eq. (\ref{hsc=0}):
\begin{eqnarray}
\left( \begin{array}{cc} \Gamma^l_0(b) & - I
\end{array} \right) \ww_{\rm inc}^{lm}
=0, \label{hsc=0Air}
\end{eqnarray}
where $\Gamma^l_0$ is the impedance tensor of the $l$th wave in the ``virtual"
air sphere. Hence, we have the relation
\begin{eqnarray}
\left( \begin{array}{cc} \Gamma^l_1(a) & - I
\end{array} \right) \Omega_{b}^{a}
=\left( \begin{array}{cc} \Gamma^l_0(b) & - I
\end{array} \right). \label{hsc=0CloakDef}
\end{eqnarray}

Impedance tensor $\Gamma^l_1$ contains material parameters
$\varepsilon^{(1)}$ and $\mu^{(1)}$ of the sphere inside the
cloaking shell. At the same time, zero scattering should be held for
arbitrary $\varepsilon^{(1)}$ and $\mu^{(1)}$ of the inner core. It implies that the partial derivative of Eq.~(\ref{hsc=0CloakDef}) with respect to $\varepsilon^{(1)}$ needs to be zero for arbitrary $\varepsilon^{(1)}$ to satisfy the zero scattering condition. Note that only the impedance tensor
$\Gamma^l_1$ contains
$\varepsilon^{(1)}$, and therefore the right-hand side of Eq.~(\ref{hsc=0CloakDef}) vanishes
after the differentiation, which results in $\frac{\partial \left(
\begin{array}{cc} \Gamma^l_1(a) & - I \end{array} \right)}{\partial
\varepsilon^{(1)}} \Omega_{b}^{a} + \left( \begin{array}{cc}
\Gamma^l_1(a) & - I \end{array} \right) \frac{\partial
\Omega_{b}^{a}}{\partial \varepsilon^{(1)} } = \left(
\begin{array}{cc} \frac{\partial \Gamma^l_1}{\partial \varepsilon^{(1)} } & 0
\end{array} \right) \Omega_{b}^{a} = \frac{\partial \Gamma^l_1}{\partial \varepsilon^{(1)} } \left(
\begin{array}{cc} I & 0 \end{array} \right) \Omega_{b}^{a} = 0$. It is now straightforward that we need the relation to be satisfied, i.e.,  $\left( \begin{array}{cc} I & 0 \end{array} \right)
\Omega_{b}^{a}=0$.

By multiplying this equation by $\Gamma^l_1(a)$
and subtracting it from Eq.~(\ref{hsc=0CloakDef}), we arrive at the
equation: $\left(
\begin{array}{cc} 0 & I \end{array} \right) \Omega_{b}^{a}=
\left( \begin{array}{cc} -\Gamma^l_0(b) & I \end{array} \right)$,
which does not contain the material parameters of the inner sphere.
Finally, we derive the evolution operator of the cloaking layer:
\begin{eqnarray}
\Omega_{b}^{a} = \left( \begin{array}{cc} 0 & 0 \\ -\Gamma^l_0(b) &
I \end{array} \right). \label{EvolOperCloak}
\end{eqnarray}

This condition defines the cloak and can be satisfied for the
specially chosen evolution operator $\Omega_{b}^{a}$ of the cloaking
shell. The evolution operator is the degenerate block matrix, whose
inverse matrix is not defined. It should be noted that relation
Eq.~(\ref{EvolOperCloak}) is independent on the material of the inner
sphere. On the other hand, the derived relation connects the wave
solutions in the cloak (evolution operator $\Omega_{b}^{a}$) with
wave solutions in the ``equivalent" homogeneous air sphere
(impedance tensor $\Gamma^l_0$). Therefore, it effectively performs
the coordinate transformation for the solutions, but not for the
material parameters as usual.

Substitution of the cloaking condition $\hh^{lm}_{\rm sc}(b)=0$ into
Eq.~(\ref{bc_final}) results in
\begin{eqnarray}
\Omega_{b}^{a} \ww_{\rm inc}^{lm} =  \left( \begin{array}{c} I \\
\Gamma^l_1(a) \end{array} \right) \! \hh^{lm}_{1}(a).
\label{bc_final1}
\end{eqnarray}
From Eq.~(\ref{EvolOperCloak}), it becomes clear that
$\hh^{lm}_{1}(a) = 0$ and $\ee^{lm}_{1}(a) \equiv
\Gamma^l_1(a)\hh^{lm}_{1}(a) = 0$. Thus one may conclude that both
electric and magnetic fields equal zero at the boundary $r=a$, and
therefore there is no field at any spatial point inside the inner
sphere. In the cloaking shell, the fields equal zero at the inner
boundary $r=a$ (owing to the continuity of the tangential fields) and
equal incident fields at the outer boundary $r=b$. Then, the field
inside the cloak must obey (see Eq.~(\ref{inside_field}))
\begin{eqnarray}
\left( \begin{array}{cc} \hh_{{\rm sh}} (\rr)
\\ \ee_{{\rm sh}}(\rr) \end{array} \right) = \sum_{l=0}^\infty \sum_{m=-l}^l
\left( \begin{array}{cc} F_{lm} & 0 \\ 0& F_{lm}
\end{array} \right)
V^l_{{\rm sh}}(r) \Omega_{b}^r \ww_{{\rm inc}}^{lm}.
\label{inside_field1}
\end{eqnarray}

\section{Nonlinear Transformation Based Spherical Cloaks}
Now, let us consider a novel class of the nonlinear transformation
based (NTB) spherical cloak, whose EM interaction can be
characterized by the proposed scattering theorem.
Fig.~\ref{Geometry_Andrey} can be regarded as the compressed space
($r$) from the original space ($r'$), i.e., an air sphere $0<r'<b$.
We propose a nonlinear transformation function \beq
r'=\frac{b^{x+1}}{(b-a)^x}\Big(1-\frac{a}{r}\Big)^x \label{r'r}\eeq
which obviously satisfies the transformation (when $r=a$, $r'=0$;
and when $r=b$, $r'=b$). The value of ``$x$" is a factor to control
the nonlinearity degree in the transformation, which can be
arbitrary from 0 to $\infty$.

Due to the invariance of Maxwell's equation under coordinate
transformations from the original space to transformed space, the
parameters ($\dy{\varepsilon}$, $\dy{\mu}$) in the shell of
Fig.~\ref{Geometry_Andrey} can be expressed in terms of those
parameters in the original space, i.e., $\dy{\varepsilon}'=1$ and
$\dy{\mu}'=1$, \beq \dy{\varepsilon}=AA^T/\det(A), \qquad
\dy{\mu}=AA^T/\det(A), \label{newParam}\eeq where $A$ is the
Jacobian matrix with elements $A_{ij}=\partial r_i/\partial r'_j$.

One can see that the proposed prescribed function Eq.~(\ref{r'r}) is
only dependent on radial position $r$. Then it is easy to find that
the Jacobian matrix is diagonal, and Eq.~(\ref{newParam}) can thus
be rewritten as \beq \dy{\varepsilon}=\dy{\mu}={\rm
diag}[\lambda^2_r,~\lambda^2_\theta,~\lambda^2_\phi]/\lambda_r\lambda_\theta\lambda_\phi={\rm
diag}[\frac{\lambda_r}{\lambda_\theta\lambda_\phi},~\frac{\lambda_\theta}{\lambda_r\lambda_\phi},~\frac{\lambda_\phi}{\lambda_r\lambda_\theta}],
\eeq where the principal stretches of the Jacobian matrix are \beq
\lambda_r=\frac{\partial r}{\partial
r'}=\frac{(b-a)^xr^{x+1}}{xab^{x+1}(r-a)^{x-1}},
~~~\lambda_\theta=\lambda_\phi=\frac{r}{r'}=\frac{(b-a)^xr^{x+1}}{b^{x+1}(r-a)^{x}}.\eeq
Finally, one can obtain the parameters of the NTB cloak ($a<r<b$) in
Fig.~\ref{Geometry_Andrey} \beq
\varepsilon_r=\mu_r&=&\frac{b^{x+1}(r-a)^{x+1}}{xa(b-a)^xr^{x+1}},
\nonumber \\
\varepsilon_\theta=\mu_\theta=\varepsilon_\phi&=&\mu_\phi=\frac{xab^{x+1}(r-a)^{x-1}}{(b-a)^xr^{x+1}}.
\eeq

Such ideal NTB spherical cloak is difficult to be fabricated in
practice. However, to some extent, it can be alleviated by dividing
the inhomogeneous cloak shell into $N$ homogeneous multilayers. The
case for cylindrical cloaks has been studied and it shows that only
several optimized layers can achieve the invisibility
\cite{Cummer_2009PRA}. Here, the optimization is out of the scope of
this paper. Our paper is to reveal some novel NTB spherical cloaks
which provides better invisibility performance than Pendry's classic
one, based on the proposed general scattering theory. The realistic
NTB cloaks can be produced using sputtering techniques, so that a
number of discrete layers should be applied over the spherical core.
To demonstrate the capability of the proposed spherical cloaks, we
present the differential cross-sections normalized by the geometric
cross-section of the cloak (see Eq.~(\ref{cross})).

In Fig. \ref{RCSvsSAnumber}, we analyze the dependence on the total
number of the layers $N$ dividing the cloaking shell. The increase
in the layer number gives rise to more accurate
approximation of the original inhomogeneous model in
Fig.~\ref{Geometry_Andrey}, and the decrease of the scattering
cross-section is expected with the increase of the number $N$. If
one uses the present scattering method with $N=50$ to divide both
Pendry's linear cloak and a specific NTB cloak at $x=1$, the forward
scattering is approximately the same, but over the whole range of
scattering angles Pendry's cloak presents better invisibility. The
following discussion will address the importance of this nonlinear
factor $x$ in beating the classic linear cloak.
\begin{figure}[htbp]
\centering\includegraphics[width=8cm]{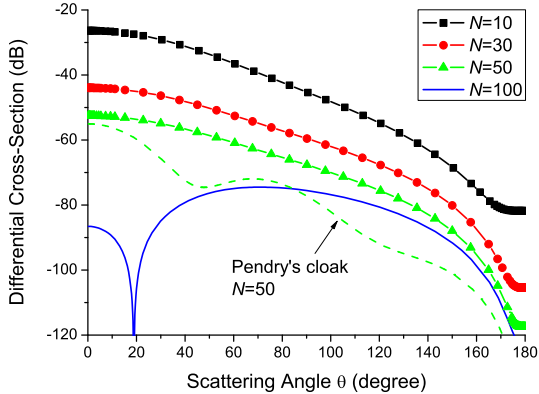}
\caption{Differential cross-section ${\rm d} \sigma/(\sigma_g
\sin\theta {\rm d} \theta)$ of the NTB spherical cloak ($x=1$) for
different number of the layers $N$ dividing the inhomogeneous
coating. Parameters: $\varepsilon^{(1)}=1.45^2$, $\mu^{(1)}=1.0$,
$k_0 a = \pi$, $k_0 b = 2 \pi$.} \label{RCSvsSAnumber}
\end{figure}

The parameter $x$ is a convenient tool to control the quality of the
NTB spherical cloak. We assume the number of homogeneous sublayers
$N=50$ for all following simulations. In Fig.~\ref{RCSvsSAx} the differential
cross-sections at different $x$ are demonstrated. If $x$ is less
than unity, the cross-section is inversely proportional to $x$,
which is not desired in the sense of invisibility. For NTB spherical
cloaks with $x<1$, the cloaking performance is degraded due to the
abrupt increase of the transverse dielectric permittivity
$\varepsilon_t$ near the inner interface $r=a$ of the clad (see
Fig.~\ref{CloakEpsilons}).

\begin{figure}[htbp]
\centering\includegraphics[width=8cm]{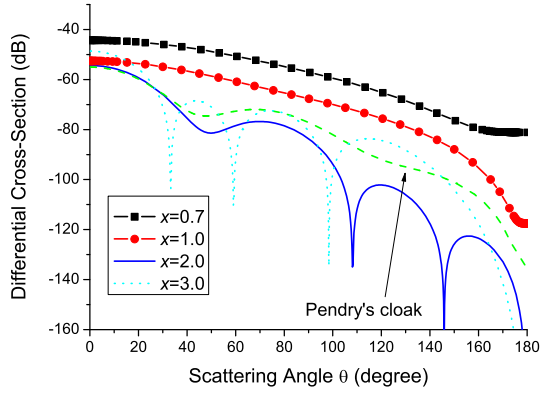}
\caption{Differential cross-section of the nonlinear cloak with
different parameters $x$. Parameters: $\varepsilon^{(1)}=1.45^2$,
$\mu^{(1)}=1.0$, $k_0 a = \pi$, $k_0 b = 2 \pi$, $N=50$.}
\label{RCSvsSAx}
\end{figure}

\begin{figure}[htbp]
\centering \includegraphics[width=8cm]{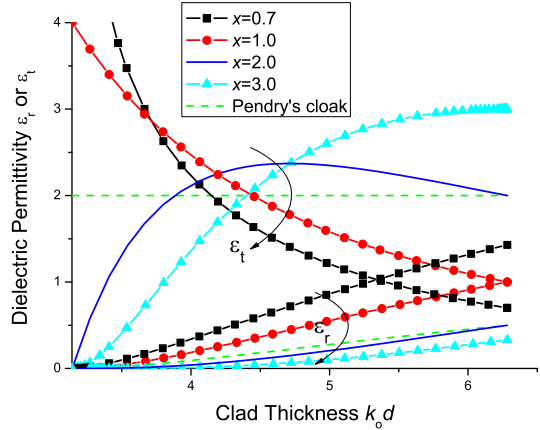} \caption{
Radial $\varepsilon_r$ and transverse $\varepsilon_t$ dielectric
permittivities of the cloaking shell with different parameters $x$.
Cloak is extended from $k_0 a = \pi$ to $k_0 b = 2 \pi$. }
\label{CloakEpsilons}
\end{figure}

The abrupt change of the material parameters is undesirable not only
for invisibility performance but also for the realization point of
view. The radial dielectric permittivity behaves in a similarly monotonic way
for all values of $x$. The dependence appears to be mainly linear
for small parameters $x$. The dependence of the transverse
dielectric permittivity is more complicated. One particular NTB cloak is
realized at the parameter $x=2$ when the transverse permittivity in
the cloak becomes non-monotonic and eventually returns to
$\varepsilon_t=2$ at $r=b$, which provides even lower cross-section
over whole observation angles than the Pendry's cloak does. If we
compare the dielectric permittivities of the proposed NTB cloak
with that of Pendry's cloak (Fig.~\ref{CloakEpsilons}), it can be noted that
the dependence of radial permittivity $\varepsilon_r$ are still quite close to each other in the cloak region. However, one may ask whether $x=2$ is the only choice or not. In Fig.~\ref{SCSvsX}, it gives the answer that in the sense of total cross section, there is a range of $x$ in which the proposed NTB spherical cloak outperforms the classic linear spherical cloak. When ``x" increases and jumps out of this optimal region, the cloaking effects compared with Pendry's cloak are degraded which can be verified by the bistatic cross-section of $x=1$ and $x=3$ in Fig.~\ref{RCSvsSAx}.
\begin{figure}[htbp]
\centering\includegraphics[width=8cm]{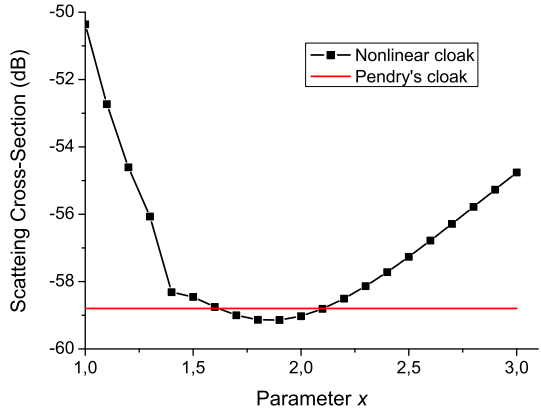}
\caption{Scattering cross-section versus $x$ for NTB spherical cloaks. Parameters:
$\varepsilon^{(1)}=1.45^2$, $\mu^{(1)}=1.0$, $k_0 a = \pi$, $k_0 b =
2 \pi$, $N=50$. The range of $x$ where SCS is lower than that of Pendry's spherical cloak is the optimal region of $x$ for desired NTB spherical cloaks.}
\label{SCSvsX}
\end{figure}

\begin{figure}[htbp]
\centering\includegraphics[width=8cm]{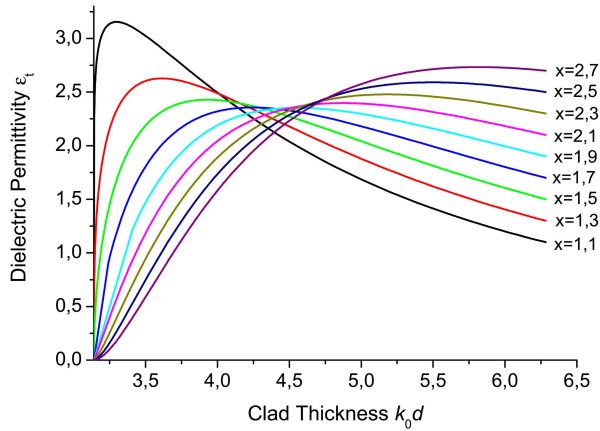}
\caption{Variance of transverse permittivity $\varepsilon_t$ along the radial direction in the region of the cloaking shell ($a<r<b$) under different values of $x$ near the optimal range as shown in Fig.~\ref{SCSvsX}. Parameters: $k_0a=\pi$ and $k_0b=2\pi$.}
\label{TrEpsilonVsThickness}
\end{figure}

Furthermore, we investigate how the transverse permittivity $\varepsilon_t$ varies near the optimal region of ``x" in Fig.~\ref{SCSvsX}. From Fig.~\ref{TrEpsilonVsThickness}, it can be observed that: 1) when $x$ is slightly above 1, the requirement of $\varepsilon_t$ near the inner boundary $r=a$ drops significantly compared to those curves whose ``x" is smaller than 1 in Fig.~\ref{SCSvsX}; 2) when ``x" becomes larger and larger, the $\varepsilon_t$ at the outer boundary $r=b$ turns to be more deviated from that value of its corresponding $\varepsilon_r$; 3) when $x$ falls into the optimal region, those curves of transverse permittivities are non-monotonic along the radial direction in the cloaking shell, and their maxima and overall values of $\varepsilon_t$ are smaller than those whose $x$ becomes further smaller or larger. These explain why there exist an optimal region for ``x" where the total scattering cross-section can be lower than Pendry's classic one. Also, it provides us another way to predict whether a certain ``x" for a NTB spherical cloak is optimal or not.

Now we continue to study the dependence of its invisibility upon the ratio $b/a$ of the particular NTB spherical cloak with $x=2$ which is discretized into $N=50$ layers. We keep the inner radius $a$ unchanged. In Fig.~\ref{RCS_SCS}(a), different ratios of $b/a$
are considered. Compared with the other three values of $b/a$, it seems that $b/a=2$ provides the best cloaking effects at nearly all angles (except for the angle at $52^\circ$) for the $x=2$ NTB cloak. Another
interesting finding is that: when $b/a>2$, the cross-section will be
higher than that of $b/a=2$ over the whole range of angles; when
$b/a \rightarrow 1$, though the angle-averaged cross-section will still be
higher but at certain angles, its cross-section could be lower than
that of $b/a=2$. From the view of total scattering, it is important to consider how the ratio $b/a$ should be selected for $x=2$ NTB cloak so as to provide improved cloaking. In Fig.~\ref{RCS_SCS}(b), one can clearly see the optimal domain of $b/a$ in which the cross-section is smaller than Pendry's spherical cloak. Certainly, all values $x$ within the desired region for the purpose of improved cloaking in Fig.~\ref{SCSvsX} have their corresponding domain
of optimal $b/a$.

\begin{figure}[htbp]
\centering \includegraphics[width=16cm]{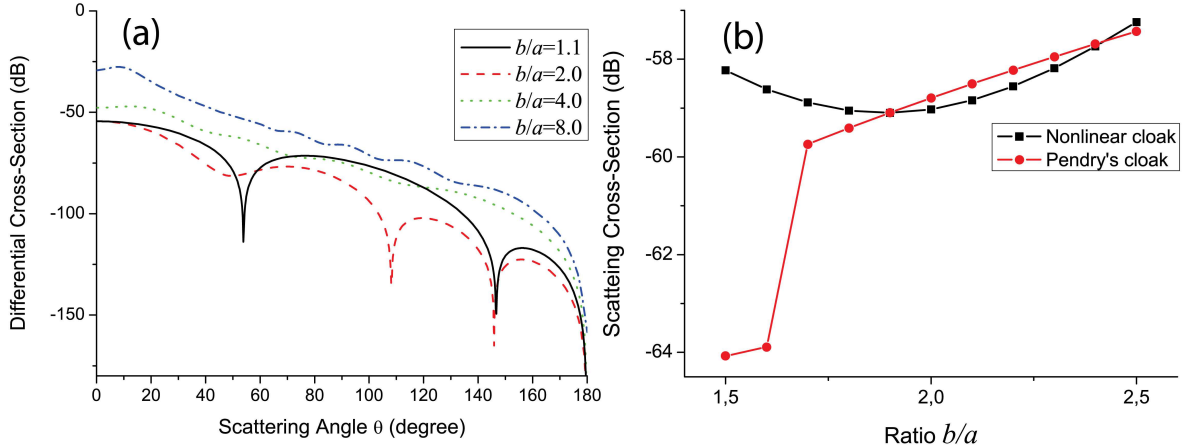} \caption{
The role of core-shell ratio $b/a$ in the cloaking improvement for $x=2$ NTB spherical cloak: (a) differential cross-section versus angle at selected ratios; (b) scattering cross-section versus ratio $b/a$. Parameters: $\varepsilon^{(1)}=1.45^2$, $\mu^{(1)}=1.0$, $k_0 a =
\pi$, and $N=50$.} \label{RCS_SCS}
\end{figure}

Let us compare the scattering diagrams for both a bare glass sphere
and a cloaked glass sphere. We take the clad in the form of the
nonlinear cloak with parameter $x=2$. From Fig.~\ref{RCSvsSAclad}, we
see that the cloaking shell noticeably reduces the scattering. Another general property of the cloaks, i.e, the exactly diminished backscattering, is also present in the figure. In what follows, we
consider their respective near-field wave interactions, which
correspond to far-field results in Fig.~\ref{RCSvsSAclad}.

\begin{figure}[htbp]
\centering \includegraphics[width=8cm]{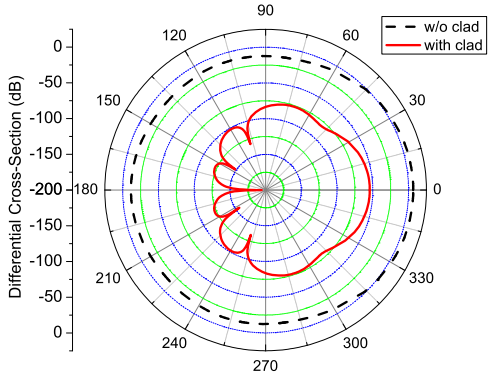} \caption{
Differential cross-section of a glass sphere with and without cloaking shell.
Parameters: $\varepsilon^{(1)}=1.45^2$, $\mu^{(1)}=1.0$, $k_0 a =
\pi$, $k_0 b = 2 \pi$, $x=2$, $N=50$.} \label{RCSvsSAclad}
\end{figure}

The near-field perturbation of the cloaked and non-cloaked glass
particles is demonstrated in Fig.~\ref{FieldCloak}. Comparing
Fig.~\ref{FieldCloak}(a) with Fig.~\ref{FieldCloak}(b), the
invisibility performance is well pronounced. In Fig.
\ref{FieldCloak}(a) the EM wave travels only through the clad and
takes near zero values in the vicinity of the inner radius $a$. The
field does not enter the glass core, therefore it does not matter
that which material is situated therein. If the cloak is less ideal
than that shown in the figure, the incident field will be scattered
by the spherical particle and will penetrate the glass core, i.e.,
the object becomes visible.

\begin{figure}[htbp]
\centering \includegraphics[width=10cm]{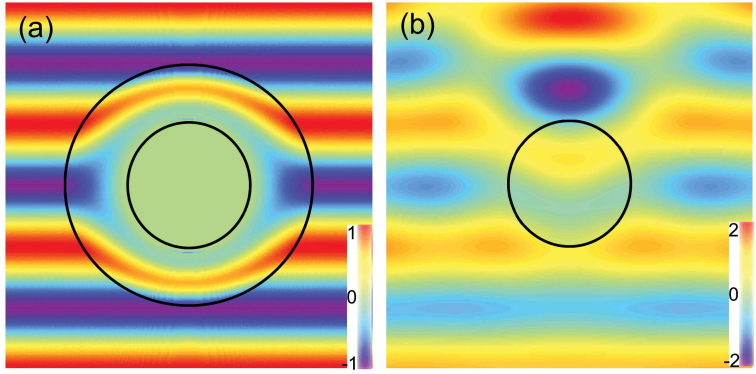} \caption{
Real part of the electric field on x-z plane scattered (a) by the
cloaking shell gathered round the glass core and (b) by the glass
core itself. Parameters: $\varepsilon^{(1)}=1.45^2$,
$\mu^{(1)}=1.0$, $k_0 a = \pi$, $k_0 b = 2 \pi$, $x=2$, $N=50$.}
\label{FieldCloak}
\end{figure}

\section{Conclusion}
We have studied the manifold of the nonlinear cloaks differing in
parameter $x$. Since there is no closed form solution for the
proposed nonlinear cloaks, an approximate model by replacing an inhomogeneous shell with homogeneous spherical layers has been numerically
analyzed, with the help of the proposed scattering algorithm for multilayered rotationally anisotropic shells. The general cloaking condition was derived from the scattering algorithm, which is in contrast to the method of coordinate transform. We have also demonstrated that better approximate spherical cloaks can be realized by properly choosing parameters $x$ and
$b/a$. In practical applications, such a class of NTB spherical cloaks can provide improved invisibility performance.
\section*{Appendix A}
\appendix
\setcounter{equation}{0}
\renewcommand{\theequation}{A-\arabic{equation}}
In order to exclude the constant vector $\hh_{1}^{lm}$, one should multiply Eq.~(\ref{bc_final}) by
$\Omega_{b}^{a}=(\Omega_{a}^{b})^{-1}$:
\begin{eqnarray}
\Omega_{b}^{a} \ww_{\rm inc}^{lm}  +  \Omega_{b}^{a} \left( \begin{array}{c} I \\
\widetilde{\Gamma}^l(b) \end{array} \right) \! \hh^{lm}_{\rm sc}(b)
= \left( \begin{array}{c} I \\
\Gamma^l_1(a) \end{array} \right) \! \hh^{lm}_{1}(a). \label{A1}
\end{eqnarray}
Then Eq.~(\ref{A1}) is further multiplied by the block matrix $\left(
\begin{array}{cc} \Gamma^l_1(a) & - I \end{array} \right)$, and the
right-hand side vanishes:
\begin{eqnarray}
\left( \begin{array}{cc} \Gamma^l_1(a) & - I \end{array} \right)
\Omega_{b}^{a} \ww_{\rm inc}^{lm}  +  \left( \begin{array}{cc} \Gamma^l_1(a)
& - I \end{array} \right)\Omega_{b}^{a} \left( \begin{array}{c} I \\
\widetilde{\Gamma}^l(b) \end{array} \right) \! \hh^{lm}_{\rm sc}(b)
= 0.
\end{eqnarray}
On the other hand, vector $\hh_{1}^{lm}$ can be
obtained in a similar way.


\newcommand{\noopsort}[1]{} \newcommand{\printfirst}[2]{#1}
  \newcommand{\singleletter}[1]{#1} \newcommand{\switchargs}[2]{#2#1}

\end{document}